# π-stacking in thiophene oligomers as the driving force for electroactive materials and devices

*Damian A. Scherlis and Nicola Marzari*

Department of Materials Science and Engineering and Institute for Soldier Nanotechnologies,

Massachusetts Institute of Technology, Cambridge MA 02139

damians@mit.edu, marzari@mit.edu

**RECEIVED DATE (to be automatically inserted after your manuscript is accepted if required according to the journal that you are submitting your paper to)**

TITLE RUNNING HEAD. π-stacking in thiophene electroactive materials

ABSTRACT. The π-stacking between aromatic oligomers has been extensively studied for many years, although the notion of exploiting this phenomenon as the driving force for molecular actuation has only recently emerged. In this work we examine with MP2 and Car-Parrinello ab-initio calculations the actuation properties of a novel class of thiophene-based materials introduced by Swager et al. (*Adv. Mat.* **2002**, *14*, 368; *J. Am. Chem. Soc.* **2003**, *125*, 1142). The chemical ingredients of the assembly, calix[4]arenes and oligothiophenes, are screened separately to characterize the actuation mechanisms and to design optimal architectures. In particular, ab-initio methods are used to study π-stacking in mixed-valence oligothiophene dimers, revealing strong interactions that can be turned on and off as a function of the electrochemical potential. We show how these interactions could be harnessed to achieve molecular actuation, and investigate the response of an active unit in real time with first-principles molecular dynamics simulations.

### Introduction

Actuator materials respond to an external signal, typically an electric field, electrochemical potential, or light, by a structural distortion that can be transduced into mechanical work. Efforts towards the development of efficient organic mechanical actuators have thrived in recent years, as these materials could provide the basis for the assembly of artificial muscles, molecular switches and motors, and other related nanodevices.[1-5]

Conducting polymers, mainly polypyrrole, polyaniline and polythiophene, are among the most successful materials to be used in the synthesis of electrochemical actuators.[1,4,6-8] The operating principle of electroactive polymer actuators is explained in terms of an exchange of ions between the electrolyte and the polymeric matrix upon oxidation/reduction. The migration of counterions inside the polymer to preserve the electrical neutrality induces a volume increase which can be reversed as a function of the electrochemical potential.[6,9,10] Such bulk mechanism, though, carries a number of limitations: the linear strain that can be achieved (usually restricted to 1-3%), the actuation frequency (limited by the diffusion of ions), and the complex nature of the bulk actuation mechanism, all of which compromise at some point the applicability of these systems in nanomechanical devices. In this context, substantial attention has been addressed to a different class of actuators in which the mechanism is driven by conformational changes at the molecular level, and does not rely on mass transport and induced swelling. Examples of this kind have been reported or proposed quite recently, where the actuation is controlled via optical,[11] electrical,[12] or chemical[13] stimuli. Some major advantages that molecular actuators could offer over bulk actuators include shorter reaction times, increased linear strain, ease of addressability, anisotropy, and the potential for implementation at the single-molecule level.

In this paper we explore a novel architecture for single-molecule actuation, whose structural, electronic and dynamical properties we thoroughly examine with static and dynamical ab-initio approaches. The design is based on the calix[4]arene-bithiophene functionalized polymers introduced by Swager et al. for applications in sensing, chemical recognition, and actuation.[14,15] Electropolymerization of the calix[4]arene bithiophene monomer allows to synthesize a material where insulating bridges are intercalated between π-conjugated oligomers. The intercalation of the calixarene moiety between quaterthiophene segments leads to a polymer with a high potential for linear elongation. We show in Figure 1a and 1b examples of the active unit and of the polymer in its ideal zigzag configuration. In this structure the calixarene plays the role of a hinge between the thiophene oligomers, which constitute the electroactive part of the material, responsible both for conductivity and for actuation. In what follows, we examine with ab-initio calculations the paradigm of molecular architectures based on switchable π-stacking interactions articulated by hinges as the driving force for microscopic and possibly macroscopic actuation. In particular, we elucidate the role of stacking interactions between oxidized thiophenes, we optimize lower rim substituents of the calix[4]erene hinges, and we offer an estimate of reaction times and dynamical response via realistic first-principles molecular dynamics simulations.

**Computational Methods**

Single-point MP2 calculations were performed employing the Gaussian 03 program.[16] MP2 interaction energies were obtained with the 6-31G(d) or the 6-311+G(d,p) basis, checking in all cases the stability of the wave functions. The basis set superposition error (BSSE) was corrected through the counterpoise method.[17]

Structural optimizations and molecular dynamics simulations were performed with the Car-Parrinello approach at the Density Functional Theory (DFT) level, as implemented in the ESPRESSO package,[18] which uses plane-wave basis sets and ultrasoft pseudopotentials to describe the electron-ion interactions.[19] The reduction in basis set size entailed by the use of ultrasoft pseudopotentials[20] in combination with the efficiency of the Car-Parrinello approach, allows for molecular dynamics studies of systems containing over a hundred atoms, as those examined here, on commodity workstations. In these plane-wave calculations the energy cutoff for the wave functions was 25 Ry, and 200 Ry for the charge density. We used the general gradient approximation (GGA) to the exchange-correlation functional in the PBE form.[21] Molecular dynamics simulations of the calixarene-quaterthiophene unit were performed in an orthorhombic cell of dimensions 28 Å x 23 Å x 15 Å, while for the full structural relaxation a box of size 28 Å x 28 Å x 15 Å was employed to minimize interactions with the periodic image. Other details on the parameters used in the calculations are provided throughout the text.

### Results and Discussion

This section is organized in four parts. The structural and electronic properties of the actuating unit and its chemical components—calixarenes and oligothiophenes—are addressed in the first three. These include: 1) performance and design of calix[4]arenes as molecular hinges; 2) effect of oxidation/reduction on the electronic structure of the polymer; 3) stacking interactions between oligothiophenes as a function of the oxidation state. Finally, in the fourth part, the dynamical response of the calixarene-thiophene architectures emerging from the previous analysis is investigated using first-principles molecular dynamics.

**Flexibility and energetics of substituted calix[4]arenes.** Calixarenes are macrocycles consisting of phenols bridged by methylene units. The calix[4]arene in particular is a versatile molecule that may adopt several conformations, depending on substituents, guest complexation, and solvent effects.[22,23] Four conformational isomers of the calix[4]arene—cone, partial cone, 1,3-alternate, and 1,2-alternate— were identified by Gutsche and Bauer twenty years ago.[24] While tetrahydroxy calix[4]arenes are trapped in the cone conformation due to intramolecular hydrogen bonding, tetra-O-alkylated and unsubstituted calix[4]arenes are free to isomerize, and several conformers have been found both experimentally and theoretically, many of them being almost equivalent in energy according e.g. to recent B3LYP calculations.[23] A calixarene intended to act as a hinge must show enough rigidity in its upper rim (the extra-annular network where the oligomers are attached) to induce a regular arrangement such as that envisioned in Figure 1, but at the same time it must offer a certain flexibility to ensure that the electroactive segments can interact without sterical or structural hindrance. In order to characterize the performance of the calix[4]arene as a flexible hinge and to assess the effect of functionalization on the lower rim we have studied the potential energy surface for the deformation of the macrocycle along the $C_a$-$C_b$ distance coordinate. $C_a$ and $C_b$ are the atoms in the upper rim connecting the benzene rings to the quaterthiophene units. The $C_a$-$C_b$ length is a crucial parameter since the energy cost to pull these centers will constrain the ability of the oligothiophene segments to approach each other. We considered here the tetra and di-hydroxy, the dimethyl, and the unsubstituted species in the 1,3-alternate or cone conformations; these are depicted in Figure 2. The potential energy surfaces for opening or closing the upper rim of these calixarenes as a function of the $C_a$-$C_b$ coordinate are shown in Figure 3. These different curves were obtained from a series of constrained relaxations at the DFT-PBE level, in which the structure was optimized at fixed $C_a$-$C_b$ distances.

The tetrahydroxy calixarene exhibits a deep minimum in the cone conformation, with the linking carbon atoms 8.6 Å apart. This is too far—as will be shown below—to enable any favorable interactions between the oligothiophenes. On the other hand, the unsubstituted macrocycle turns out to be extremely floppy, with a negligible barrier to drive the distance between terminal carbon atoms $C_a$ and $C_b$ from 5 to 10 Å (the sudden increase in energy beyond 10.5 Å is due to the stretching of the molecule—at that distance the benzene rings become parallel at an angle of 180 degrees).

Finally, for the case of the dimethyl calixarene, the full geometry relaxation leads to a 1,3-alternate configuration as the global minimum, 8.6 kcal/mol more stable than the cone conformer. In this 1,3-alternate isomer steric interactions between the methyl groups in the lower rim prevent the flipping of the substituted aromatic rings, which in turn affects the separation between the terminal atoms $C_a$ and $C_b$ (see Figure 2). Such flipping could lead to an opening of the calixarene hinge that would undermine the desirable zigzag design. The corresponding energy profile, with a minimum at a distance of 5.2 Å, makes the dimethyl calix[4]arene a convenient scaffold to link the oligothiophene segments without hindering their mutual interactions but preserving at the same time the optimal arrangement for molecular actuation. The replacement of two hydroxyl groups in the lower rim to obtain a dimethyl calixarene, though, has been seldom reported in the literature (for a review on the subject, see Ref. 25). Alternatively, the dihydroxy derivative, which is more amenable from the synthetic point of view, exhibits a similar behaviour, albeit with a smaller energy difference between the cone and the 1,3-alternate isomers.

Table 1 summarizes the geometrical and energetic parameters for these different derivatives, including as an additional case the dimethoxylated calix[4]arene. In this derivative the $OCH_3$ groups constrain the flipping of the benzene rings, imposing a high barrier between the two almost isoenergetic isomers.

**Electronic structure of the polymer: effect of oxidation.** As will be discussed below, the electrochemical potential can be used to control the interactions between thiophene oligomers. The effects of the electrochemical potential on the electronic structure of the polymer were investigated by computing the charge density difference between the neutral and the oxidized states. We plot in Figure 4 the isodensity surfaces for this difference, obtained from PBE-GGA calculations in periodic boundary conditions on a unit cell of size 32 Å x 26.42 Å x 16 Å and containing 160 atoms. The positive and negative lobes—corresponding to the density difference between the neutral and the charged polymer[26]—illustrate the regions most affected upon oxidation. It can be seen that the effect of oxidation is almost completely confined to the oligothiophene, whereas the electronic density of the calixarene remains practically unaffected. This result reinforces the notion of the calixarene as an electronically inert moiety with a defined structural function, and the quaterthiophene as the electroactive element of the assembly.

**Interactions between thiophene oligomers controlled via the electrochemical potential.** The considerations outlined above allow us to focus on the interactions between oligothiophenes disregarding initially the presence of the calix[4]arene, and assuming that their stacking properties will be essentially the same within the assembly. Thiophene oligomers and polymers are among the most important and promising organic conductive materials, and their ability to form π-stacks, fundamental in the understanding of structural and transport properties, has been studied and discussed intensively in recent years.[27-32] Parallel thiophenes in the neutral state interact weakly through attractive van der Waals forces, in general not larger than 3 kcal/mol.[32] In the case of oxidized oligothiophenes, though, π-π attractive terms compete with the Coulombic repulsion. In a previous work we have applied highly-correlated techniques to the case of oxidized bithiophenes,[31] finding that for charged dimers of bithiophenes the electrostatic repulsion totally overwhelms the π-π binding between radical cations, leading to unfavorable interactions of the order of 50 kcal/mol even at distances as large as 5 Å. The presence of a solvent dramatically affects the repulsion of these charged dimers, leading to reversible formation of stable dimers both experimentally[27,28] and theoretically.[31] This stabilization in solution may be interpreted recalling that a polarizable dielectric promotes concentration of charge in smaller cavities, thus favoring a doubly charged bound dimer with respect to the dissociated singly charged constituents. Three major factors can thus be identified that determine the stacking of oxidized thiophene oligomers: the π-π interactions driven by the favorable mixing of semioccupied orbitals, the Coulombic repulsion, and the polarization effect of a solvent.

At the redox potential the neutral form of the oligothiophene will coexist with the oxidized species, leading to possible mixed-valence states. Dimers between a radical cation (or anion) and its parent molecule are well known in various organic species in solution and in crystals.[33,34] However, to the best of our knowledge, neither experimental nor theoretical studies have been conducted to establish the existence of mixed-valence stacking in thiophene oligomers or to provide an estimation of the thermodynamical stability. From the theoretical side, this problem is fairly challenging since it cannot be tackled with conventional electronic structure approaches as Hartree-Fock or DFT. First, the importance of electronic correlation in the description of π-stacking[31,32,35,36] renders the Hartree-Fock method unsatisfactory to treat this kind of systems. Second, self-interaction errors in DFT induce a delocalization of the unpaired electron in the radical dimer, yielding an unphysical charge of -0.5 e on each monomer. As a consequence, post-Hartree-Fock, highly-correlated schemes are needed to provide reliable results. Here we employ the MP2 method, which has proven successful in the past to investigate the stacking of neutral[35,36] and charged[31] aromatic compounds.

The lower panel in Figure 5 depicts the MP2 interaction energy at the 6-311+G(d,p) level, of a mixed valence bithiophene dimer as a function of the inter-planar distance between the two monomers, which are facing each other in parallel planes. Interestingly, the calculations show that a neutral and a singly charged bithiophene will interact strongly in vacuum, resulting in the formation of singly charged dimers with binding energies of around 12 kcal/mol and intermolecular distances of 3.5 Å. This trend is general for oligothiophenes of different lengths: MP2/6-311G(d,p) geometry optimizations on a singly-charged monothiophene dimer yield a bound state with a binding energy of 23.4 kcal/mol (Figure 6), whereas MP2/6-31G(d) single-point calculations on a mixed valence, parallel terthiophene dimer gave an interaction of 11.4 kcal/mol at a separation of 3.6 Å (this energy value should be regarded as a lower limit since the geometry of the dimer was not relaxed). For comparison, the upper panel of Figure 5 shows the strong repulsive behavior found in vacuum for doubly charged dimers of bithiophene cations.

This tendency to dimerize clearly apparent in the potential energy surface, is also revealed in the MP2 single determinant picture by the formation of molecular orbitals of bonding (and antibonding) character in the mixed valence dimer. The original orbital pattern of the monomers is still recognizable in the dimer, distinctly hybridized to contribute to localized and delocalized orbitals extending in between the molecular planes. The high spin contamination typically found in systems requiring a multi-determinant description was not observed in this case.

**Molecular dynamics simulations.** The dramatic change in the intermolecular forces between oligothiophenes, from a bound mixed valence dimer to a highly repulsive potential in the doubly-charged state (Figure 5), could provide the driving force for a thiophene-based molecular actuator to change shape in response to oxidation or reduction. In an optimal design the $C_a$-$C_b$ distance in the calixarene should be tuned to the stacking distance in the dimer, and at the same time should present a potential energy surface shallow enough to allow flexibility. The results shown in Figure 3 and Table 1 point to the dihydroxy and the dimethyl substituted calixarenes as some that, in their most stable (alternate) configuration, best fulfill these requirements. In order to study the dynamical response to oxidation, we decided to perform first-principles Car-Parrinello molecular dynamics simulations on a model system of the actuating unit involving two quaterthiophene segments attached to the dihydroxylated calixarene. The quaterthiophene oligomers were connected to the upper rim through the $C_a/C_b$ positions shown in Figure 2. The initial geometry corresponded to a constrained relaxation performed on the neutral system, since the treatment of the singly charged state is not accurate in DFT for the reasons given above. The only constraint in this relaxation was the inter-planar distance between the quaterthiophene arms, fixed to 3.5 Å as found in the mixed-valence dimer.[37] The strong nature of the bond for the mixed-valence state, ascertained by the MP2 calculations, together with the flexibility of the calixarene, assure us that this initial structure is a good approximation to the singly charged bound configuration of the complex. DFT-GGA on the other hand provides a reliable quantitative description of the doubly charged system.[31] We present in Figure 7 our results for the time evolution of the potential energy and the angle between the two quaterthiophenes. At time zero, two electrons were removed and the system was allowed to evolve freely. A rapid drop in the energy of at least 10 kcal/mol takes place within the first few tenths of a picosecond. First, the system releases the repulsive energy by an in-plane lateral displacement of the initially overlapping oligomers. This is followed by an expansion of the calixarene upper rim in which the oligothiophene arms move further from each other keeping their parallel configuration. Throughout this early reorganization the oligomers remain parallel, explaining why the strong decrease in the potential energy is not immediately correlated with an increase in the opening angle. As the result of the electrostatic repulsion the system continues its expansion to reach a maximum aperture of nearly 65° in 2.5 ps. By the end of our simulation time the system is still highly perturbed. We expect that a longer simulation would ultimately lead to equilibration at a smaller angle, since a full geometry optimization performed in a cell of dimensions 28 Å x 28 Å x 15 Å yielded an opening of 43.5° for the relaxed structure.

Figure 8 presents the same information for a second model structure, in which the quaterthiophene arms are linked to the two non-methylated rings of a dimethyl calixarene. The behaviour in this system, aside from small quantitative differences, is comparable to the one exhibited by the previous one, confirming the robustness of the actuation mechanism driven by the quaterthiophene interactions. The structure reaches a maximum opening after 4 ps and then oscillates around 35°. In this case the simulation was started from a less contracted structure, corresponding to a geometry relaxed in the neutral state without constraints. This may partly explain both the longer times involved in achieving maximum expansion and the smaller drop in potential energy (the plotted energies are relative to the energy at 0 ps). Moreover, the geometry minimization yielded a final angle of 37.2°, which can account for the smaller expansion observed in Figure 8. In both systems these deformations represent a linear spatial expansion of over 100%, with responses to oxidation that are broadly similar. Further engineering of the calix[4]arene hinge or functionalization of the oligothiophenes could be envisioned, for synthetic, processing or mechanical purposes. This actuation paradigm could also be combined with other supramolecular connectors different from calix[4]arenes; experimental and computational studies in this direction are under way.

**Conclusions**

In this work, we have studied the electronic, structural and dynamical properties of a novel class of calixarene-thiophene based materials designed for single molecule actuation. In particular, the calculation of the potential energy surfaces of several calix[4]arene derivatives allowed us to characterize the effects of substitutions on the lower rim, and to optimize the choice of the calixarene to be used as a structural hinge. Furthermore, we showed that singly charged thiophene oligomer dimers exhibit strong π-stacking in vacuum. We proposed a way to harness these interactions, driven by changes in an applied electrochemical potential, in a design which combines calixarenes as flexible hinges and oligothiophenes as active elements in a molecular actuator with potential for extensive deformation. First-principles molecular dynamics simulations on this optimized calixarene-quaterthiophene unit provided an estimate of the overall performance and operation times.

ACKNOWLEDGMENT. We thank T. Swager, I. Hunter and P. Anquetil for sharing their expertise on the calixarene-oligothiophene compounds. This research was supported by the Institute of Soldier Nanotechnologies, contract DAAD-19-02-D0002 with the U.S. Army Research Office.

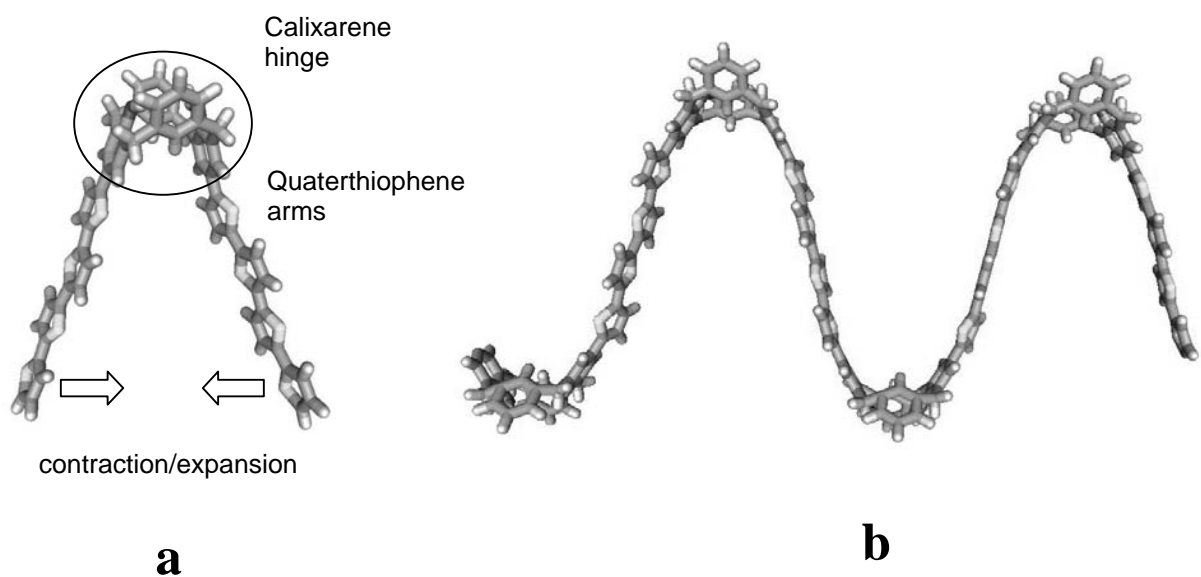

**Figure 1.** Structure of a calix[4]arene-thiophene molecular actuator: one actuating unit (a) and a polymerized assembly (b).

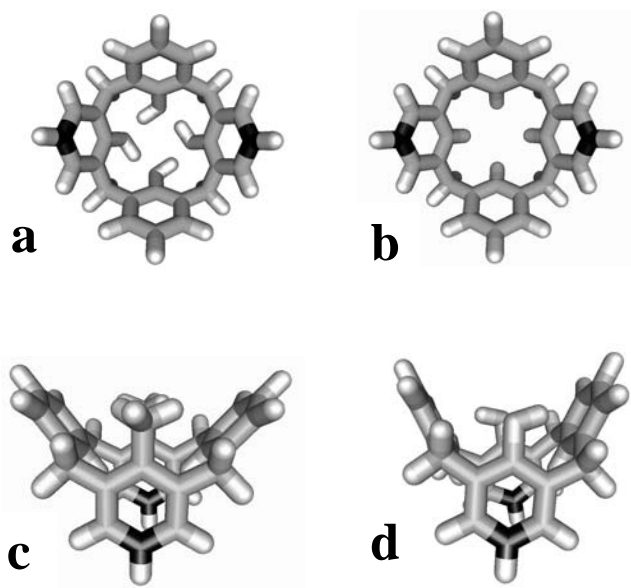

**Figure 2.** Substituted calix[4]arenes in their minimum energy conformations: (a) tetrahydroxylated cone; (b) unsubstituted cone; (c) dimethylated 1,3-alternate; (d) dihydroxylated 1,3-alternate. The linking atoms $C_a$ and $C_b$ connecting the upper rim to the quaterthiophenes are indicated in black.

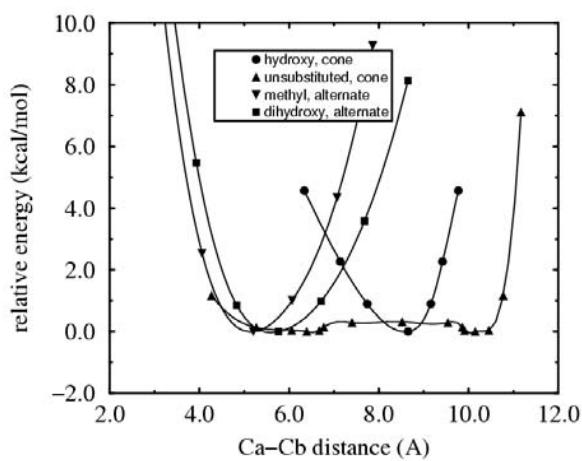

**Figure 3.** Potential energy surfaces for the deformation of substituted calixarenes as a function of the distance between the linking atoms $C_a$ and $C_b$ (see Figure 2).

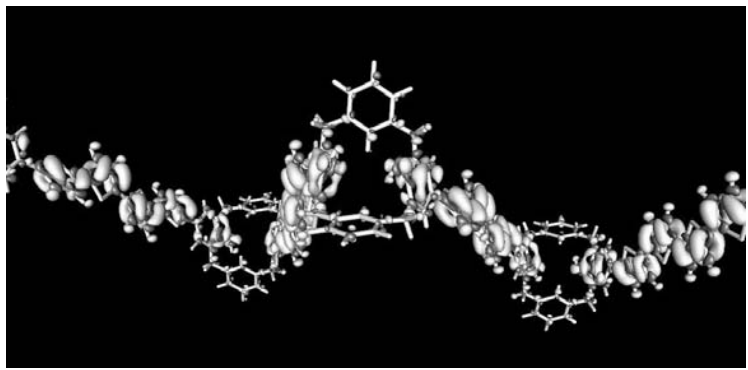

**Figure 4.** Electronic density difference between the neutral and the oxidized states of the calixarene-thiophene polymer shown in Figure 1.

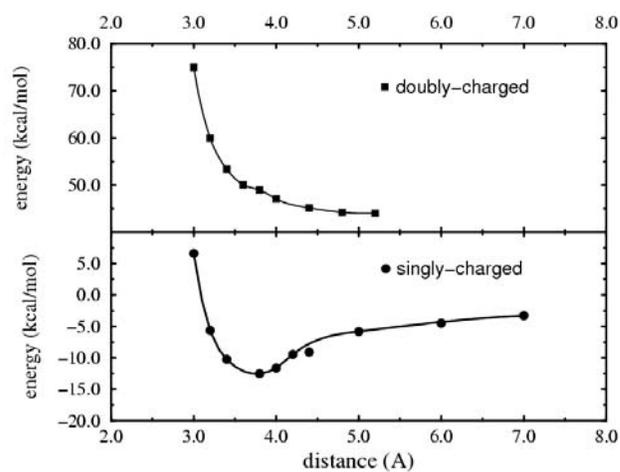

**Figure 5.** MP2 interaction energies as a function of the inter-planar distance for a bithiophene dimer in its doubly charged (upper panel) or singly charged state (lower panel).

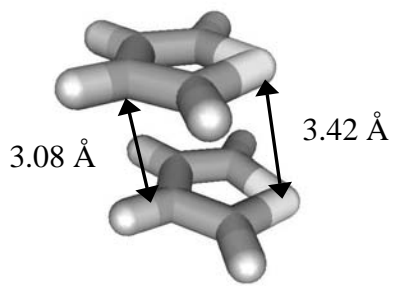

**Figure 6.** Optimized geometry for a singly charged monothiophene dimer at the MP2 level.

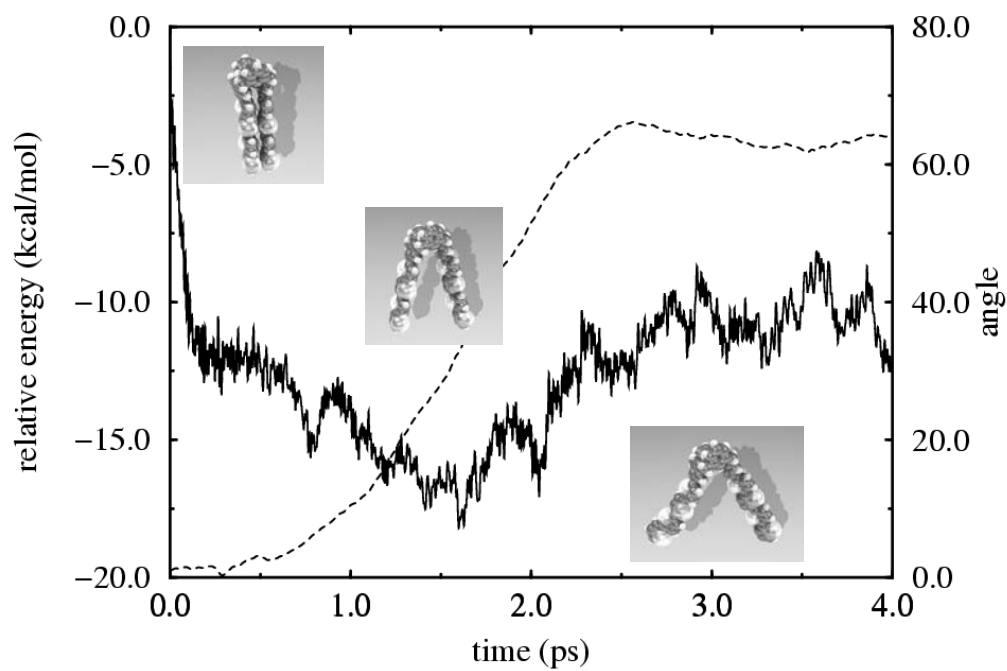

**Figure 7.** Time evolution of the potential energy (solid line) and the angular opening (dashed line) of a dihydroxy calixarene-quaterthiophene unit upon oxidation. The molecular models correspond to snapshots at 0, 1.5 and 3 picoseconds.

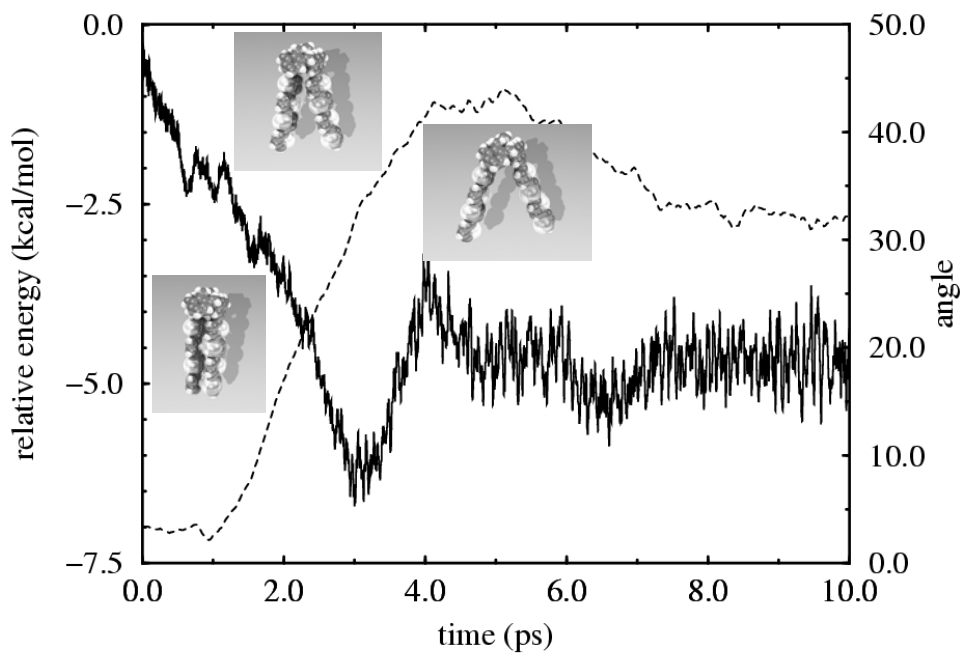

**Figure 8.** Time evolution of the potential energy (solid line) and the angular opening (dashed line) of a dimethyl calixarene-quaterthiophene unit upon oxidation. The molecular models correspond to snapshots at 0, 2.5 and 5 picoseconds.

**Table 1.** Relative energies and $C_a$-$C_b$ distances for the 1,3-alternate and cone conformations of substituted calix[4]arene.

| derivative | rel. energy (kcal/mol) | | $C_a$-$C_b$ (Å) | |
|---|---|---|---|---|
| | alternate | cone | alternate | cone |
| hydroxylated | [a] | 0.0 | [a] | 8.6 |
| dimethylated | 0.0 | 8.6 | 5.2 | 7.9 |
| dimethoxylated | 0.4 | 0.0 | 5.4 | 9.4 |
| dihydroxylated | 0.0 | 1.7 | 5.8 | 10.2 |

[a]No minimum found.

SYNOPSIS TOC.

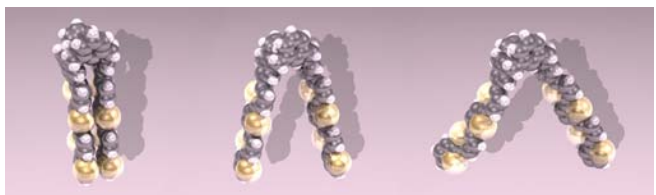